\documentclass{aa}  
\usepackage{graphicx}
\usepackage{natbib}
\usepackage{amssymb}
\usepackage{amstext}
\usepackage{latexsym} 
\usepackage[sumlimits]{amsmath}
\newcommand{\Ha}{H$\alpha$}			
\newcommand{\NII}{[N{\sc ii}]}			
\newcommand{\HII}{H{\sc ii}}			
\newcommand{\HI}{H{\sc i}}			

\usepackage{txfonts}
\begin{document}
   \title{Star-forming galaxies in low-redshift clusters: Effects of
environment on the concentration of star formation \thanks{
Based on observations made with the Nordic Optical Telescope,
operated on the island of La Palma jointly by Denmark, Finland, Iceland,
Norway, and Sweden, in the Spanish Observatorio del Roque de los 
Muchachos of the Instituto de Astrof\'isica de Canarias; and with 
the Jacobus Kapteyn Telescope, which was operated on the island of La Palma
by the Isaac Newton Group in the Spanish Observatorio del Roque de los
Muchachos of the Instituto de Astrof\'isica de Canarias. 
}}
   \titlerunning{Environment and concentration of star formation}

	\author{C.F.~Bretherton\inst{1,2}
	 \and C.~Moss\inst{1}\thanks{Deceased 12th May 2010}
	 \and P.A.~James\inst{1}}

   \institute{Astrophysics Research Institute, 
	Liverpool John Moores University, 
	Birkenhead CH41 1LD
	\and Carter Observatory, Wellington, New Zealand
}

   \date{Received ; accepted }

 
  \abstract
   {}
   {We attempt to determine the dominant processes acting on star-forming
disk galaxies as a result of the cluster environment by studying the 
normalised rates and radial distributions of 
star formation in galaxies within low-redshift clusters.}
   {We develop indicators of different processes based on the radial
concentrations of $R$-band and H$\alpha$ light within each of the galaxies
studied. The tests are applied to galaxies in 
each of 3 environments - cluster, supercluster (outside the cluster
virial radius) and field. We develop new diagnostic diagrams combining 
star-formation rate and spatial distribution information to differentiate 
between stripping of outer disk gas, general gas depletion, nuclear starbursts
and galaxy-wide enhancement of star formation.
}
   {Hubble type classifications of cluster galaxies are found to correlate only weakly with their concentration indices, whereas this correlation is strong for non-cluster 
populations of disk galaxies.  We identify a population of early-type disk galaxies in the cluster
population with both enhanced and centrally-concentrated star formation compared to their field counterparts.
The enhanced cluster galaxies frequently show evidence of disturbance.
A small but non-negligible population of cluster galaxies with truncation 
of star formation in their outer disks is also found.
}
   {}

   \keywords{galaxies: clusters: general - galaxies: evolution - 
galaxies: interactions - galaxies: structure - galaxies: stellar content -
stars: formation}

   \maketitle
%

\section{Introduction}

This is the third paper in a series studying the star formation (SF) 
properties of a carefully-selected sample of disk  galaxies lying
within 8 local clusters.  The galaxies are sufficiently large and
nearby that their properties can be studied in some detail, even with
ground-based imaging.  The overall aims of the programme are to identify
characteristic and distinctive modes of star formation that specifically occur
within cluster galaxies, and to use these properties to select between the 
many effects that have been proposed to affect cluster galaxies.  These
effects are predicted to occur in 
different locations within clusters, and over widely varying timescales.

The first paper in the series \citep{tho08} discussed the galaxy
sample selection (227 galaxies in 8 low-redshift clusters) and the
\Ha\ and $R$-band imaging data that were taken to enable the analysis
of SF and structural properties of these galaxies.  This paper also
looked at completeness of the sample as a function of \Ha\ flux,
equivalent width (EW) and surface brightness, and a catalogue was
presented of derived global parameters for all 227 galaxies.  The
second paper \citep{bre10} contained a detailed comparison of the SF
properties of the cluster galaxies, as represented by their global
\Ha\ EW values, with the same properties of both a `supercluster'
sample (galaxies lying outside the virial radius of any cluster, but
in survey fields centred on clusters), and a field galaxy sample from
the \Ha\ Galaxy Survey \citep[henceforth \Ha GS; N.~Shane PhD thesis 2002;][]{jam04}.  A substantial population
of cluster galaxies with enhanced \Ha\ emission relative to 
field galaxies of the same morphological type was identified; the enhanced galaxies frequently
show signs of disturbance. Marginal evidence was found for a higher
velocity dispersion of the enhanced galaxies, possibly indicative of
an infalling population.

In the present paper, a further analysis of these cluster galaxies
will be presented, with the major aim being to determine the nature of
the enhancement of line emission found in many of the cluster
galaxies: is this primarily a disk phenomenon, or concentrated in the
central bulge regions, or is the increased SF a global phenomenon,
enhancing the SF activity throughout the affected galaxies?  How does
the cluster environment affect the development of disks and bulges?

The extensive literature on the observed properties of cluster galaxies,
and on the physical processes likely to be acting, precludes any attempt
at a comprehensive discussion of the literature on these topics.  Some of the
most relevant papers were discussed briefly in \citet{bre10}; this will not be
repeated here, and much more comprehensive accounts are available in review
articles, e.g. \citet{bos06}.

This paper concentrates on the radial distribution of SF activity in
disk galaxies, which has been addressed in several previous studies,
using a variety of SF tracers and looking at galaxies in different
environments.  One of the earliest of these was \citet{hod83}, who
presented radially-binned counts of numbers of \HII\ regions within 37
spiral galaxies, from narrow-band \Ha\ imaging.  For this sample of
predominantly field galaxies, three characteristic radial patterns
were found - exponentially-declining, `doughnut' distributions with a
peak at intermediate radius, and oscillating distributions with two or
more radial peaks.  The types of radial distribution were fund to
correlate both with the presence of bars and with Hubble type.
\citet{ryd94} compared the \Ha\ , $V$- \& $I$- band exponential
scale-lengths of a sample of 34 southern field spirals.  They found
the \Ha\ emission, tracing ongoing SF, to have much larger
scale-lengths than light in either of the continuum bandpasses, where
the latter presumably trace older stellar populations.

\citet{koo98}, in the first paper of an important study of Virgo cluster 
galaxies which will form a benchmark reference for the present work, used
\Ha\ imaging to illustrate that the overall suppression of SF in cluster 
galaxies occurs preferentially in outer disk regions.  They also made
the claim, reinforced in their later papers \citep[e.g.][]{koo04a} that the
resulting centrally-concentrated SF could cause the erroneous 
mis-classification of stripped late-type spirals as earlier-types.
In \citet{koo06} they studied $R$-band and \Ha\ scale lengths for their
Virgo cluster galaxies, finding further evidence for truncation of the
\Ha\ emission that was not seen in a comparison sample of field spirals.

\citet{dal01} presented an analysis of rotation curves and \Ha\ line
emission from 510 cluster spiral galaxies.  One of the parameters they
studied was the radial extent of the \Ha\ emission, normalised by
galaxy radii determined from $I$-band imaging, as a function of
morphological type and position within the host cluster. They concluded
that line emission is seen to extend to larger galactocentric radius
in galaxies that lie further from cluster centres, which is most
simply interpreted in terms of gas stripping from outer disk regions
of galaxies lying close to cluster cores.

\citet{hat04} studied the distribution of \Ha\ luminosity in 22 bright
star-forming galaxies, and found evidence for centrally-concentrated
emission, particularly among the disturbed galaxies.  \citet{ben07}
used mid-infrared 24~$\mu$m emission to map SF morphologies within
galaxies from the Spitzer Infrared Nearby Galaxies Survey, finding
typically compact and symmetric emission in early-type spirals, and
extended and asymmetric emission in late-types.

Finally, \citet{mun07} compared GALEX UV emission profiles (assumed to
be a tracer of recent SF) with 2MASS near-infrared data (tracing the
older stellar population) for 161 predominantly field galaxies. They
found that the average specific SF rate rises with increasing distance
from galaxy centres, consistent with the findings of \citet{ryd94},
and of \citet{koo06} for their field comparison sample.

\section[]{Data}

The samples of galaxies that are analysed in the present paper are
discussed fully in \cite{tho08}, but given the importance of the
selection criteria for the interpretation of any results, a brief
overview will be given here.  All apart from the field comparison
sample are subsets of galaxies detected as emitters of \Ha\ and/or
\NII\ lines by the Objective Prism Survey \citep[henceforth
  OPS;][]{mos00,mos05}.  The OPS looked at 8 low-redshift galaxy
clusters, with numbers 262, 347, 400, 426, 569, 779, 1367 and 1656 in
the catalogue of \citet{abe58}, and characterised the line emission
properties of 727 galaxies from the CGCG catalogue \citep[][and
  references therein]{zwi68}.  Two main subsamples were chosen.  The
first was all spiral galaxies with Hubble types between Sa and Sc
inclusive (the `Sa - Sc sample') in all clusters except Abell~262 and
Abell~347, giving a complete sample of galaxies in this type range for
the remaining 6 clusters.  The second is all galaxies identified as
emission-line galaxies in all 8 clusters of the OPS (the `ELG'
sample). The completeness of the ELG sample with respect to \Ha\ flux,
EW and surface brightness is discussed in full in
\citet{tho08}, where it is shown that the sample becomes significantly
incomplete below an \Ha\ $+$ \NII\ EW of 2~nm, below a flux of
3.2$\times 10^{-17}$~W~m$^{-2}$ and below a surface brightness of
4.0$\times 10^{-20}$~W~m$^{-2}$~arcsec$^{-2}$.

Galaxies from the OPS are designated as `cluster' galaxies if they lie
within the virial radius \citep[defined in][]{bre10}, and as
`supercluster' galaxies if they lie outside this radius.  A matched
comparison sample of `field' galaxies was selected from the \Ha GS
survey \citep{jam04}, again as described in \citet{bre10}.
The `cluster' and `supercluster' galaxies were further subdivided
according to the strength of their \Ha\ $+$ \NII\ emission
in \citet{tho08}; `enhanced' and `non-enhanced' subsets were defined
according to whether galaxies lie within or outside the 2-$\sigma$ EW limits 
derived from the field galaxies of the same type.

As explained in paper 2, many of the cluster galaxies had no
catalogued morphological classifications.  These cases were addressed
by classifications performed by Mark Whittle, as explained in
\citet{mos00}.  All classifications used are on the de Vaucouleurs
system \citep{dev59,dev74}, and intercomparisons were made with the
UGC \citep{nil73} and RC3 \citep{dev91} catalogues where possible.
Particular note was made of any signs of tidal disturbance.  For
 consistency with earlier papers in this series we used `by eye'
methods to identify galaxies with disturbed morphologies.
'Disturbed' classification were given to galaxies if their R-band
images showed signs of strong tidal features and corresponding
asymmetry, tidal tails or significant warping of the disk plane.

The observational data discussed here are $R$-band and
\Ha\ narrow-band imaging, taken with either the Jacobus Kapteyn
Telescope (JKT) or the Nordic Optical Telescope, both of which are on
the island of La Palma, in various runs between 1994 and 2005. Typical
seeing for the images is 1.5~arcsec, corresponding to a spatial resolution 
of $\sim$0.6~kpc for galaxies in these clusters, at distances of 75 - 100~Mpc.
The galaxies studied have effective radii of $\sim$10~kpc, 
and thus our images give
good resolution of their morphological and star-formation properties. 

\section[]{Concentration indices as a measure of SF distribution}
\label{ch:conc_ind}

A major advantage of photometric \Ha\ imaging is that it allows star formation to be mapped across whole galaxies. As well as providing global star formation data, the distribution of star formation as a function of radial distance from the galaxy centre can also be investigated, and individual regions identified. 

A first method to compare the distribution of star formation between galaxies is to examine the concentration of \Ha\ emission across each galaxy. Comparing this with the distribution of continuum emission across the same galaxy provides an insight into the location of star forming regions relative to the underlying, older stellar population. A straightforward approach to this is to use concentration indices to investigate the relative compactness of star formation in sample galaxies. Such indices provide a quantitative determination of the radial distribution of light in specified passbands that is both useful in itself, and also provides a more objective indicator of galaxy type than is provided by, for example, Hubble classifications.

\subsection{$R$-band concentration parameter}

$R$-band central light concentrations are related to bulge-to-disk or bulge-to-total light ratios, and to the Hubble type of spiral galaxies. They are also independent of the models assumed for galaxy components, and, therefore, provide an objective measure of the radial distribution of galaxy light. Following the methods of \citet{koo01} and \citet{koo04b}, the $R$-band central light concentration parameter we use here is defined as
\begin{equation}
C_{30} = \frac{F_{R}(0.3r_{24})}{F_{R}(r_{24})}.
\end{equation}

This is based on a similar parameter used by \citet{abr94}.  In this definition, $F_{R}(r_{24})$ is the total $R$-band flux measured within the 24 magnitudes per arcsec$^{2}$ isophote, i.e. within $r_{24}$, and $F_{R}(0.3r_{24})$ is the flux measured within 0.3$r_{24}$.  As many of the galaxies are disk systems with substantial inclinations to the line of sight, elliptical apertures matched in shape to the outer isophotes were used in calculating all concentration indices. We estimate errors on the $C_{30}$ values to be at the 5 per cent level, in agreement with \citet{koo98}.

In \citet{jam09}, we performed a comparison of $C_{30}$ and two other concentration indices, one of which was the `classical' index of \citet{dev77} based on radii containing percentiles of the total flux, and the other based on Petrosian radii \citep{pet76}.  The indices were calculated from the $R$-band light distributions of galaxies in the \Ha GS field sample, and all were found to correlate well with Hubble type, such that earlier types have indices showing more centrally-concentrated $R$-band light distributions, as expected.  Of the three, the $C_{30}$ index was found to give marginally the best correlation with Hubble type, and hence it was adopted for use in the present study.

\begin{figure}
\vspace{-30mm}
\includegraphics[width=87mm]{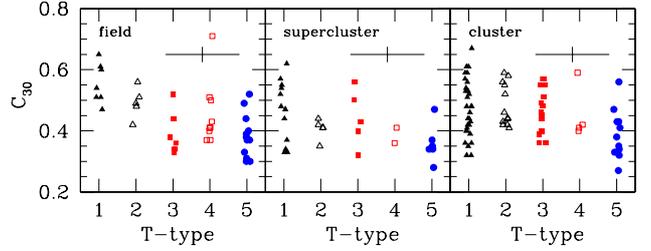}
\vspace{-28mm}
\caption{$R$-band concentration index $C_{30}$ as a function of galaxy
$T$-type showing a good correlation for the field sample, but with significant
scatter for all types. For cluster galaxies, the correlation is weaker overall, and a number of the early-type cluster galaxies have low $R$-band light concentrations. Typical errors applicable to all points are indicated by the bars
in the upper-right corner of each frame.}
\label{fig:C30dept}
\end{figure}

Figure \ref{fig:C30dept} shows $C_{30}$ plotted against morphological type for field, supercluster and cluster Sa--Sc samples. A Kendall rank test shows substantial correlation ($\tau = -0.425$) at nearly the 4$\sigma$ level for the field sample, but with significant scatter in $C_{30}$ for any given type. For the supercluster objects, this correlation is still fairly strong ($\tau = -0.34$), although the result is less significant (2.7$\sigma$). The cluster $R$-band concentration indices, however, although still weakly correlated with morphological type ($\tau = -0.18$, 2.3$\sigma$) have a much wider range than for the field objects, particularly for early type spirals, with half of all cluster galaxies of type Sa having $C_{30}$ values below those for any field Sa galaxy. A K-S test gives a probability of $\sim$8\% that the field and cluster Sa indices are drawn from the same distribution. Table \ref{tbl:c30stats} summarises the same information as Fig. \ref{fig:C30dept}, giving means, standard errors on means, medians and standard deviations for the $C_{30}$ distributions as a function of galaxy type and environment.

\begin{table*}
\begin{center}
\begin{small}
\begin{tabular}{|c||c|c|c|c|c||c|}
\hline
Type & Sa & Sab & Sb & Sbc & Sc & Environment\cr
\hline
\hline
$<C_{30}>$(Std. Err.) & 0.56(0.02) & 0.49(0.02) & 0.39(0.03) & 0.46(0.04) & 0.38(0.02) & Field \cr
$\sigma_{C_{30}}$ & 0.065 & 0.051 & 0.069 & 0.107 & 0.071 & \cr
Median$_{C_{30}}$ & 0.54 & 0.49 & 0.36 & 0.41 & 0.37 & \cr
\hline
$<C_{30}>$(Std. Err.) & 0.46(0.03) & 0.41(0.02) & 0.44(0.04) & 0.39(0.03) & 0.36(0.02) & Supercluster \cr
$\sigma_{C_{30}}$ & 0.097 & 0.034 & 0.092 & 0.035 & 0.062 & \cr
Median$_{C_{30}}$ & 0.47 & 0.41 & 0.43 & 0.39 & 0.35 & \cr
\hline
$<C_{30}>$(Std. Err.) & 0.47(0.02) & 0.49(0.02) & 0.47(0.02) & 0.46(0.05) & 0.39(0.02) & Cluster \cr
$\sigma_{C_{30}}$ & 0.094 & 0.069 & 0.068 & 0.090 & 0.081 & \cr
Median$_{C_{30}}$ & 0.47 & 0.45 & 0.45 & 0.42 & 0.38 & \cr
\hline
\end{tabular}
\caption{Statistical properties of the distributions of $C_{30}$ values for 
field, supercluster, and cluster Sa-Sc galaxies.}
\label{tbl:c30stats}
\end{small}
\end{center}
\end{table*}

\citet{koo98,koo04b} find similar results for Virgo cluster spirals and suggest that the Hubble sequence may not provide a meaningful classification system for cluster galaxies. In particular they find that a number of Virgo galaxies classified as Sa may in fact be strongly stripped low to intermediate concentration (later type) galaxies with truncated star forming disks. \citet{koo04b} suggest that these misleading classifications may explain much of the larger mean \HI\ deficiency of Virgo Sa galaxies compared to those of types Sb and Sc, and the systematic differences in kinematics and spatial distributions of early and late type cluster spirals. These authors also suggest that this effect may at least partially explain the excess of galaxies classified as early type spirals in nearby clusters, relative to lower-density environments.

For the current field sample, and the isolated galaxies observed by Koopmann and Kenney, $C_{30}$ is found to correlate well with Hubble type. Indeed, a good correlation between Hubble type and central concentration is seen for field galaxies in general (e.g. \citealt{ken85,con03}). \citet{koo04b} bin their data into four $C_{30}$ ranges that roughly correspond to isolated S0, Sa, Sb, and Sc galaxies. These ranges are shown in Table \ref{tbl:c30range}.

It is clear that $C_{30}$ must be correlated with bulge-to-disk ratio, although this is in a non-linear fashion (\citealt{gra01}). \citet{koo01} give mean [$C_{30}$:B/D] pairs as [0.3:0.1], [0.4:0.3], [0.5:0.8], and [0.6:1.0], although each value has a scatter of at least $\pm$0.1.

\begin{table*}
\begin{center}
\begin{tabular}[h]{|c||c|c|c|c|}
\hline
\textbf{Type} & S0 & Sa & Sb & Sc \cr
\hline
\textbf{$\textbf{C}_{\textbf{30}}$ range} & 0.61--0.72 & 0.51--0.60 & 0.38--0.50 & 0.00--0.37 \cr
\hline
\end{tabular}
\caption[$C_{30}$ ranges corresponding to isolated S0, Sa, Sb, and Sc galaxies]{$C_{30}$ ranges corresponding to isolated S0, Sa, Sb, and Sc galaxies taken from \citet{koo04b}.}
\label{tbl:c30range}
\end{center}
\end{table*}

\begin{figure}
\vspace{-30mm}
\includegraphics[width=87mm]{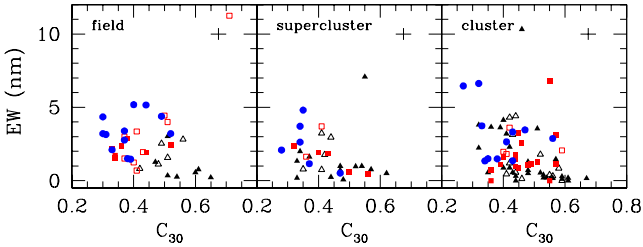}
\vspace{-28mm}
\caption{EW vs. $C_{30}$ for field, cluster and supercluster Sa--Sc samples. Symbols correspond to galaxy type as shown in Fig. 1.  All samples show at least a
moderate trend for galaxies with high concentration indices to have weaker
\Ha\ emission. Typical error bars are shown in the upper-right corner of each frame.}
\label{fig:ewdepc30}
\end{figure}

Figure \ref{fig:ewdepc30} shows EW plotted against $C_{30}$. This shows evidence of a correlation, with late-type, low concentration galaxies tending to have higher EW values, and therefore higher star formation rates normalised by their continuum luminosities. This dependence of normalised star formation rate on Hubble type has been discussed in many papers in the literature, including \citet{ken83a}, \citet{ken83b}, \citet{deh94}, \citet{ryd94} and \citet{you96},  and is reviewed by \citet{ken98}. A Kendall rank test, however, shows only a moderate correlation between $C_{30}$ and EW, at the 2$\sigma$ level, for the field Sa--Sc sample, although this is stronger and more significant (3.2$\sigma$) when the full field sample is considered. The supercluster Sa--Sc sample shows a similar relationship, significant at the 2.4$\sigma$ level. Despite the lack of correlation between type and $C_{30}$ for the cluster data, however, the relationship between $C_{30}$ and EW is strongest ($\tau = -0.31$) for the cluster Sa--Sc sample, and is significant at nearly the 4$\sigma$ level. 

\subsection{\Ha\ concentration parameter}

The \Ha\ concentration index used here is defined in a similar way
to that for $R$-band light, as follows:
\begin{equation}
C_{\rm{H}\alpha} = \frac{F_{\rm{H}\alpha}(0.3r_{24})}{F_{\rm{H}\alpha}}
\end{equation}

where $F_{\rm{H}\alpha}$ is the total \Ha\ flux measured within the $r_{24}$ isophote, and $F_{\rm{H}\alpha}(0.3r_{24})$ is the flux measured within 0.3$r_{24}$ as above. This value will be relatively low for objects with disk-dominated star formation. For galaxies with truncated star forming disks or enhanced star formation in circumnuclear regions, the \Ha\ emission will be more centrally concentrated, leading to higher values of $C_{\rm{H}\alpha}$. Errors on $C_{\rm{H}\alpha}$ are somewhat larger than those for $C_{30}$, due to the `clumpy' nature of H$\alpha$ emission.  We estimate typical errors on $C_{\rm{H}\alpha}$ values to be 10 per cent, from an analysis of the size of the steps in the H$\alpha$ growth curves of galaxies in our sample, which contrast with the smoother growth curves in the $R$-band light.

\begin{figure}
\vspace{-30mm}
\includegraphics[width=87mm]{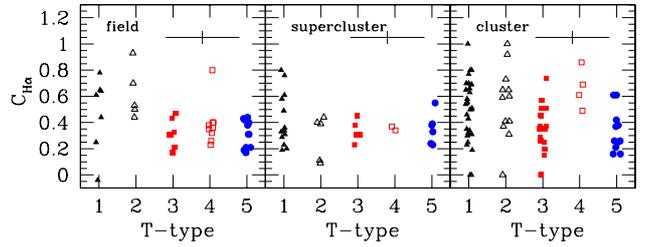}
\vspace{-28mm}
\caption{Distribution of $C_{\rm{H}\alpha}$ with $T$-type. Five cluster galaxies with no detected \Ha\ emission have been excluded from the plot. Typical error bars are shown in the upper-right corner of each frame.}
\label{fig:chat}
\end{figure}

Figure \ref{fig:chat} shows $C_{\rm{H}\alpha}$ plotted against $T$-type for the three Sa--Sc samples. There is some scatter in $C_{\rm{H}\alpha}$ across all types in the field, supercluster and cluster samples, and this scatter is found to be particularly large for the cluster sample. There is a moderate correlation ($\tau =-$0.3, 2.9$\sigma$) between $C_{\rm{H}\alpha}$ and $T$-type for the field sample, but only a low correlation for the cluster data. A number of cluster objects appear to have relatively high values of $C_{\rm{H}\alpha}$ compared to the field galaxies of the same type.  Table \ref{tbl:cHastats} summarises the statistical information on the $C_{H\alpha}$ value distributions.

\begin{table*}
\begin{center}
\begin{small}
\begin{tabular}{|c||c|c|c|c|c||c|}
\hline
Type & Sa & Sab & Sb & Sbc & Sc & Environment\cr
\hline
\hline
$<C_{H\alpha}>$(Std. Err.) & 0.48(0.11) & 0.62(0.09) & 0.32(0.04) & 0.39(0.06) & 0.31(0.03) & Field \cr
$\sigma_{C_{H\alpha}}$ & 0.29 & 0.20 & 0.11 & 0.17 & 0.11 & \cr
Median$_{C_{H\alpha}}$ & 0.61 & 0.53 & 0.31 & 0.36 & 0.31 & \cr
\hline
$<C_{H\alpha}>$(Std. Err.) & 0.42(0.06) & 0.29(0.08) & 0.34(0.04) & 0.36(0.02) & 0.35(0.05) & Supercluster \cr
$\sigma_{C_{H\alpha}}$ & 0.21 & 0.17 & 0.08 & 0.02 & 0.12 & \cr
Median$_{C_{H\alpha}}$ & 0.34 & 0.39 & 0.31 & 0.36 & 0.36 & \cr
\hline
$<C_{H\alpha}>$(Std. Err.) & 0.54(0.04) & 0.60(0.07) & 0.39(0.04) & 0.66(0.08) & 0.34(0.05) & Cluster \cr
$\sigma_{C_{H\alpha}}$ & 0.20 & 0.22 & 0.16 & 0.16 & 0.16 & \cr
Median$_{C_{H\alpha}}$ & 0.55 & 0.60 & 0.37 & 0.65 & 0.26 & \cr
\hline
\end{tabular}
\caption{Statistical properties of the distributions of $C_{H\alpha}$ values for 
field, supercluster, and cluster Sa-Sc galaxies.}
\label{tbl:cHastats}
\end{small}
\end{center}
\end{table*}

\begin{figure}
\vspace{-30mm}
\includegraphics[width=87mm]{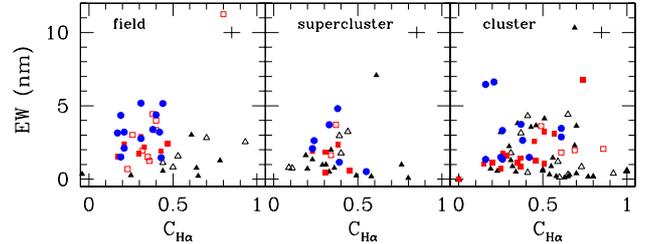}
\vspace{-28mm}
\caption{Dependence of EW on $C_{\rm{H}\alpha}$ for Sa--Sc objects. Little or no correlation is seen for field, supercluster or cluster samples.Typical error bars are shown in the upper-right corner of each frame.}
\label{fig:ewdepcha}
\end{figure}

EW is plotted against $C_{\rm{H}\alpha}$ in Fig. \ref{fig:ewdepcha}. There does appear to be some correlation between $C_{\rm{H}\alpha}$, $T$-type and EW for the field sample, but a Kendall rank test shows very little dependence of EW on $C_{\rm{H}\alpha}$ over all types. There is essentially no correlation ($\tau =$ 0.08) between $C_{\rm{H}\alpha}$ and EW for the cluster galaxies, with a 92\% probability that a correlation of this strength could be obtained by chance.

The UV-bright, Sbc starburst galaxy, NGC 3310, stands out in the field sample as having both high and relatively concentrated \Ha\ emission for its type. This same object, which has a complex peculiar morphology thought to result from a recent minor merger, also has a high value of $C_{30}$ (Figs. \ref{fig:C30dept} and \ref{fig:ewdepc30}), particularly when compared to other later-type spirals. 

\section{Comparison of H$\alpha$ and $R$-band concentration indices}
\label{sec:cvc}

\subsection{The Sa--Sc sample}

In this section we will compare the values of $C_{\rm{H}\alpha}$ found for galaxies in different environments, by plotting them against the $C_{30}$ values found for the same galaxies.  Here, $C_{30}$ should be seen as a more objective proxy for galaxy type. This analysis will be done for all galaxy subsamples (cluster, supercluster, field, disturbed, undisturbed, Sa - Sc and ELG).

\begin{figure}
\vspace{-5mm}
\includegraphics[width=110mm]{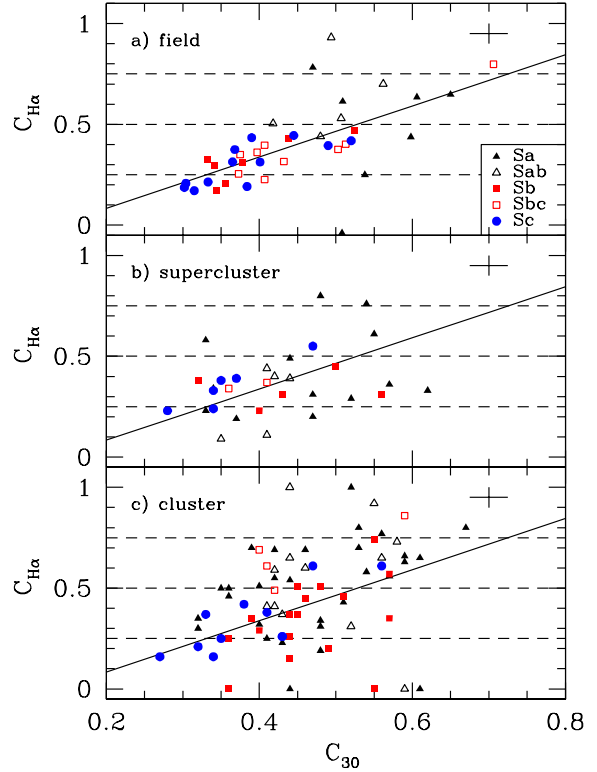}
\vspace{-5mm}
\caption{\Ha\ concentration index, $C_{\rm{H}\alpha}$, plotted against $R$-band central light concentration, $C_{30}$, for field, supercluster and cluster Sa--Sc galaxies. The dashed lines show the values of $C_{\rm{H}\alpha}$ at which an object has three-quarters, half and one-quarter (top to bottom) of its \Ha\ emission within 0.3$r_{24}$. The solid line has equation  $C_{\rm{H}\alpha}=1.269C_{30}-0.170$ and is the best fit line to the field data. Cluster Sa--Sc spirals tend to have more concentrated \Ha\ emission, and hence star formation, relative to $R$-band concentration, than their field counterparts. The cluster galaxies lying at $C_{\rm{H}\alpha}$ = 0 are those for which no \Ha\ was detected. Typical error bars are shown in the upper-right corner of each frame.}
\label{fig:cvc_all}
\end{figure}

\begin{figure}
\vspace{-30mm}
\includegraphics[width=110mm]{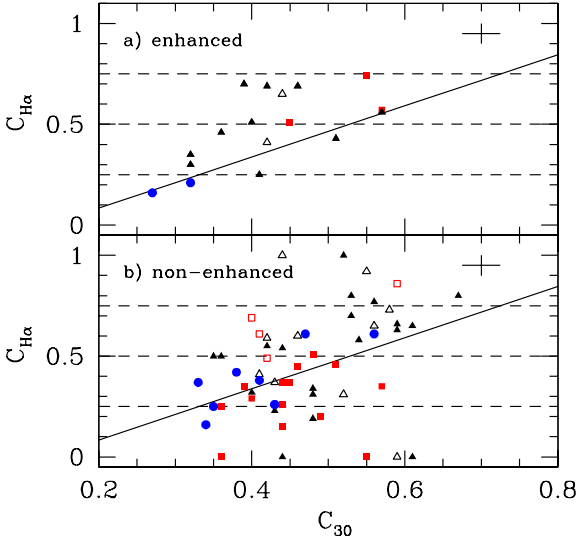}
\vspace{-5mm}
\caption{$C_{\rm{H}\alpha}$ vs. $C_{30}$ for cluster Sa--Sc data, split into enhanced and non-enhanced objects. The solid and dashed lines and point types are the same as used in Fig. \ref{fig:cvc_all}. Typical error bars are shown in the upper-right corner of both frames.}
\label{fig:cvc_clen}
\end{figure}

\begin{figure}
\vspace{-30mm}
\includegraphics[width=110mm]{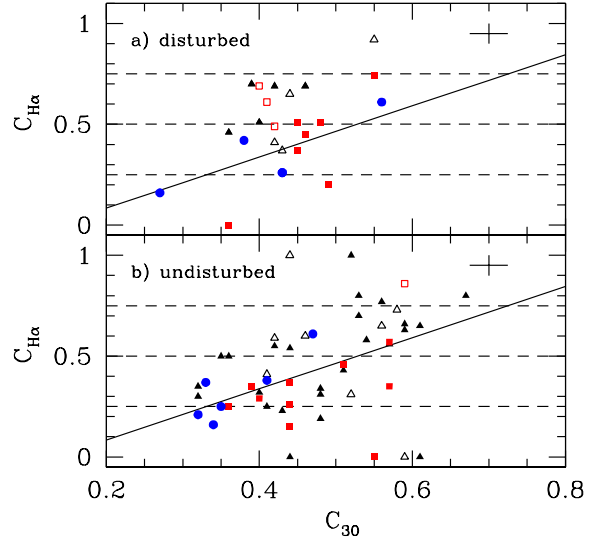}
\vspace{-5mm}
\caption{$C_{\rm{H}\alpha}$ plotted against $C_{30}$ for cluster Sa--Sc galaxies, here split into disturbed and undisturbed samples.Typical error bars are shown in the upper-right corner of both frames.}
\label{fig:cvc_cldist}
\end{figure}

Figure \ref{fig:cvc_all} shows \Ha\ vs. $R$-band concentration indices for field, supercluster and cluster Sa--Sc galaxies. Here the dashed horizontal lines show the $C_{\rm{H}\alpha}$ values for galaxies containing three-quarters, half and one-quarter of their \Ha\ emission within 0.3$r_{24}$. The points lying at $C_{\rm{H}\alpha}=0$ in the cluster plot are for the five galaxies with no detectable \Ha\ emission. These are included on the plot for completeness, but they are excluded from the subsequent analysis. The fractions of points lying in each of the ranges of $C_{\rm{H}\alpha}$ are listed in Table \ref{tbl:cHarange}.  This table summarises the data plotted in Figs. \ref{fig:cvc_all} - \ref{fig:cvc_cldist}. 

\begin{table*}
\begin{center}
\begin{small}
\begin{tabular}{|c||c|c|c|c|c|c|c|}
\hline
$C_{H\alpha}$   & Field & Supercluster & Cluster & Cluster & Cluster & Cluster & Cluster \cr
range       &       &              & all     & enh.    & nonenh. & dist.   & undist. \cr  
\hline
0.00 - 0.25 &  0.25 & 0.26         & 0.14    & 0.18    & 0.17    & 0.09    & 0.17    \cr
0.25 - 0.50 &  0.53 & 0.58         & 0.39    & 0.29    & 0.39    & 0.39    & 0.39    \cr
0.50 - 0.75 &  0.15 & 0.10         & 0.36    & 0.53    & 0.30    & 0.48    & 0.30    \cr
0.75 - 1.00 &  0.08 & 0.06         & 0.10    & 0.00    & 0.13    & 0.04    & 0.13    \cr
\hline
\end{tabular}
\caption{Fraction of $C_{H\alpha}$ values lying in the 4 ranges demarked by the
dashed lines in Figs. \ref{fig:cvc_all} - \ref{fig:cvc_cldist}.}
\label{tbl:cHarange}
\end{small}
\end{center}
\end{table*}

\citet{koo04b} plot a similar diagram for their Virgo cluster and isolated galaxy samples. These authors suggest that Virgo cluster galaxies typically have most of their star formation within 0.3$r_{24}$ ($C_{\rm{H}\alpha}<0.5$), whilst isolated spirals tend to have more star formation beyond 0.3 $r_{24}$ ($C_{\rm{H}\alpha}>0.5$). They also find a paucity of low-$C_{30}$ isolated galaxies with $C_{\rm{H}\alpha}>0.65$, compared to 25\% of Virgo objects.

A similar result can be seen in Fig. \ref{fig:cvc_all} with 49\% of cluster galaxies having $C_{\rm{H}\alpha}$ greater than 0.5 (i.e. more than half of their star formation within 0.3$r_{24}$) compared to 16\% for the supercluster and 22.5\% for the field samples. A K-S test shows that the distributions of $C_{\rm{H}\alpha}$ for field and cluster Sa--Sc galaxies are different at $>$98\% significance. 

The solid lines in Fig. \ref{fig:cvc_all} show the best fit to the field data with equation $C_{\rm{H}\alpha}=1.269C_{30}-0.170$ and a root-mean-square (rms) scatter of 0.15. The supercluster data also agree well with this relationship, with a similar rms scatter of 0.16. As would be expected, approximately half of the field/supercluster data points fall above the line and half below. The cluster data, however, are not well fit by the line, with over 65\% of points falling above the field line. Assuming a binomial distribution with an equal chance of a point appearing above or below the line, the probability of finding 45 out of 69 cluster galaxies lying above the best fit field line is less than 0.004.  It is important to note that this result is independent of the adopted classifications for these galaxies, and depends only on their $R$-band and \Ha\ light distributions.

Figure \ref{fig:cvc_clen} shows the same cluster data, but this time split into enhanced and non-enhanced objects. For the non-enhanced sample there is still a tendency towards more concentrated \Ha\ emission, with $\sim$62\% galaxies falling above the line, a similar fraction to the overall population. For the 17 cluster galaxies identified as having enhanced emission, a similar or stronger trend is indicated, with 13 ($>$76\%) having more concentrated \Ha\ emission, relative to $C_{30}$, than the average field object. The probability of this occurring by chance is $<$2\%. Considering only early-type spirals, some 83\% (10 out of 12) of the enhanced Sa--Sab galaxies appear above the line. 

In \citet{bre10} it was shown that EW correlates well with disturbed morphology, such that the most disturbed galaxies tend to have higher EW values, and therefore stronger star formation per unit luminosity. The cluster Sa--Sc data are therefore plotted again (Fig. \ref{fig:cvc_cldist}), split into disturbed and undisturbed objects. As with the enhanced objects, the disturbed galaxies tend towards higher \Ha\ concentrations (71\% above best fit field line) particularly once again for early-type spirals, with seven of the eight disturbed Sa--Sb cluster galaxies having more concentrated \Ha\ emission. Approximately 60\% of undisturbed galaxies, however, also lie above the line.

These results show that disturbed galaxies have more centrally concentrated star formation, often with relatively high EW values. The tendency for enhanced galaxies also to have higher values of $C_{\rm{H}\alpha}$, and the correlation between enhancement and tidal disturbance, suggests that a number of objects may be experiencing an increase of star formation in circumnuclear regions, triggered by some form of tidal interaction. 

It is interesting to note, however, that all seven Sa--Sc galaxies with more than three-quarters of their \Ha\ emission within $r_{24}$, and several other objects with relatively concentrated \Ha\ emission, are in the non-enhanced sample. Similarly, a number of objects lying above the field line show few signs of tidal disturbance. Coupled with the fact that there is no significant correlation between $C_{\rm{H}\alpha}$ and EW (see Fig. \ref{fig:ewdepcha}) this suggests that concentrated \Ha\ emission may be linked not only to an increase of star formation within 0.3$r_{24}$ for some objects, but also to a decrease of star formation in the outer regions of a number of galaxies. \citet{koo04b,koo04a} suggest that the cause of such a decrease is the truncation of the star forming disk due to processes such as ram-pressure stripping. The relative importance of increased central SF and tidal truncation of outer disks will be analysed further in Sect.~\ref{sec:efr}.

\subsection{The ELG sample}

\begin{figure}
\vspace{-5mm}
\includegraphics[width=110mm]{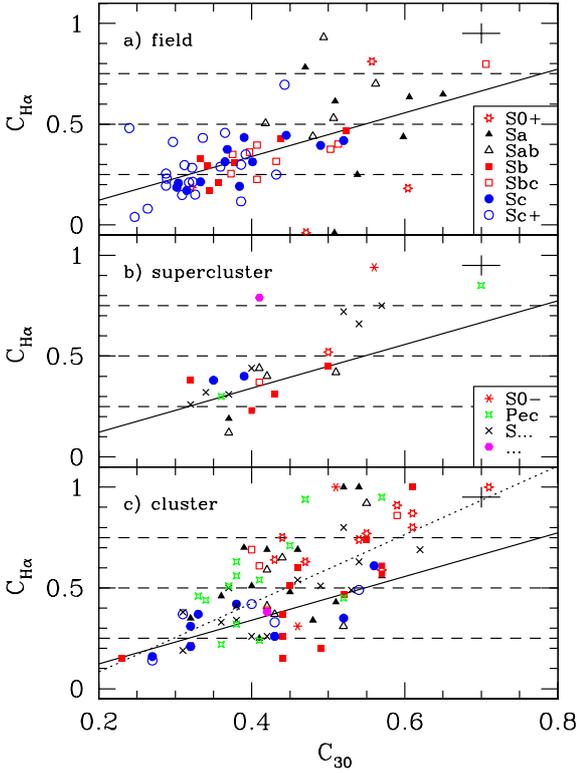}
\vspace{-5mm}
\caption{\Ha\ concentration index, $C_{\rm{H}\alpha}$, plotted against $R$-band concentration index, $C_{30}$, for the full field comparison sample and the cluster and supercluster ELG objects. Here the solid line is the best fit to the full field data and has equation  $C_{\rm{H}\alpha}=1.085C_{30}-0.094$. The dotted line in the lower panel shows an alternative best fit line, with equation  $C_{\rm{H}\alpha}=1.702c_{30}-0.257$, calculated from only the cluster ELG data. This shows that the relationship between $C_{30}$ and $C_{\rm{H}\alpha}$ is noticeably steeper for the cluster sample than it is for the field. Typical error bars are shown in the upper-right corner of each frame.}
\label{fig:cvc_elgall}
\end{figure}

\begin{figure}
\vspace{-30mm}
\includegraphics[width=110mm]{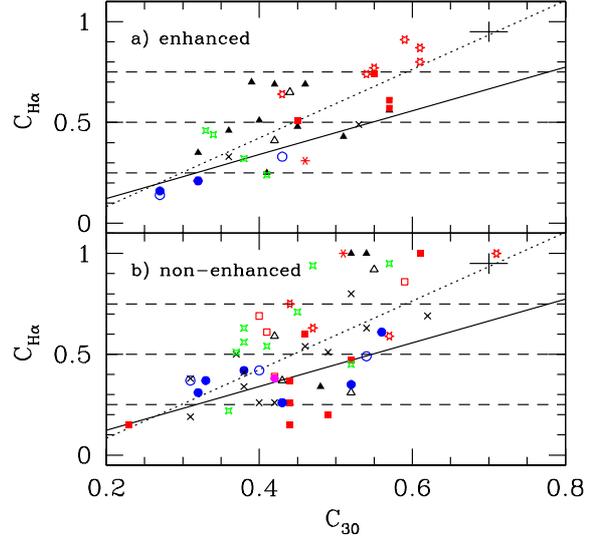}
\vspace{-5mm}
\caption{Comparison of \Ha\ and $R$-band concentration indices for cluster ELG data, split into enhanced and non-enhanced objects.Typical error bars are shown in the upper-right corner of both frames.}
\label{fig:cvc_elgen}
\end{figure}

\begin{figure}
\vspace{-30mm}
\includegraphics[width=110mm]{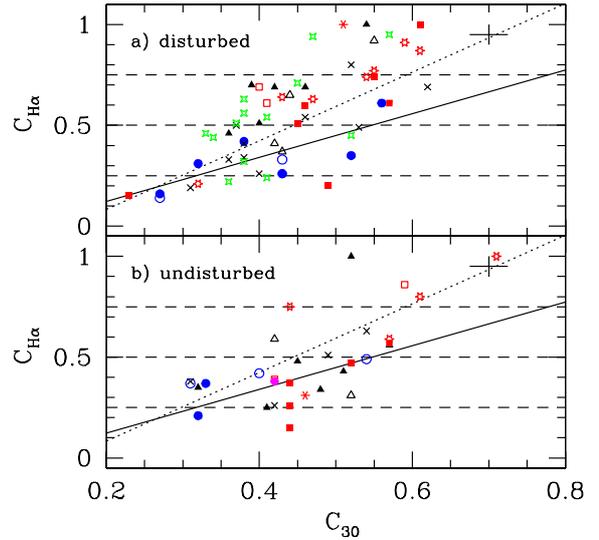}
\vspace{-5mm}
\caption{$C_{\rm{H}\alpha}$ vs. $C_{30}$ for disturbed and undisturbed cluster objects. The majority of disturbed ELGs lie above both the field and cluster best fit lines, whilst the undisturbed objects are broadly consistent with the field data.Typical error bars are shown in the upper-right corner of both frames.}
\label{fig:cvc_elgdist}
\end{figure}

We can check the robustness of our concentration index results to possible selection effects by applying the same analysis to the ELG sample, which is selected purely on the detection of H$\alpha$ line emission in the OPS and is completely independent of, for example, potentially subjective galaxy classifications. Similar plots to Figs. \ref{fig:cvc_all}--\ref{fig:cvc_cldist} were constructed for the ELG sample, and are shown in Figs. \ref{fig:cvc_elgall}, \ref{fig:cvc_elgen} and \ref{fig:cvc_elgdist}. Here the dashed lines are the same as for the previous plots, but the solid line now shows the best fit to the full field sample, including both earlier S0/a and later Sc--Irr objects. This fit is similar to that for the Sa--Sc sample with equation  $C_{\rm{H}\alpha}=1.085C_{30}-0.094$, and again provides a good fit to the field data, with a similar rms scatter of 0.16.

The cluster and supercluster samples include some galaxies of earlier morphological type than the field sample, as well as a number of peculiar objects and galaxies of uncertain spiral type. However, the best fit field line is independent of type and a single relation between $C_{30}$ and $C_{\rm{H}\alpha}$ fits well across the full range of $R$-band concentrations. The additional cluster and supercluster objects can therefore be compared to the same trend when considering their \Ha\ and continuum concentration indices. Also shown in yellow are the two galaxies for which no type information is available.

Figure \ref{fig:cvc_elgall} shows data for the full field S0/a--Irr and the cluster and supercluster ELG samples. The supercluster ELG sample contains only 25 objects; these do, nevertheless, appear reasonably well fit by the field line at lower values of $C_{30}$ ($\lesssim$0.5). There are several objects, however, particularly with higher $C_{30}$ values, which fall well above the line. 

The cluster late-type Sc and Scd--Irr galaxies also appear well fit by the field line, but, as is seen in the Sa--Sc sample, the earlier type cluster objects tend to have relatively high values of $C_{\rm{H}\alpha}$. Overall, some 70\% of cluster ELGs have \Ha\ concentrations that result in their lying above the field line. Assuming a binomial distribution, the probability of finding 63 of 90 objects lying above the line (when expecting only half) is around $6 \times 10^{-5}$. For galaxies of type Sab and earlier, the fraction above the line rises to nearly 80\% (probability $<$ 0.001). Also noticeable is the large excess of peculiar galaxies (77\%) with more concentrated \Ha\ emission. This is consistent with the suggestion of \citet{mos06} that these objects may be the results of recent merger events.

The ELG sample is biased towards higher \Ha\ surface brightness, based on analysis of the completeness of the Objective Prism Survey from which it is selected, and objects with compact emission may be expected to have higher surface brightnesses in central regions. A Kendall rank test on the complete Sa--Sc sample, however, shows no significant correlation between $C_{\rm{H}\alpha}$ and total \Ha\ surface brightness, and only a very low correlation between \Ha\ concentration index and the \Ha\ surface brightness within 0.3$r_{24}$. This is certainly not sufficient to explain the large excess of cluster ELG objects lying above the best fit field line.

Taking the cluster ELG data alone, a new best fit line is found to have equation  $C_{\rm{H}\alpha}=1.702C_{30}-0.257$. This reduces the rms scatter for the cluster data from 0.21 (around the field line) to 0.16, and is plotted as the dotted line in the bottom section of Fig. \ref{fig:cvc_elgall} and subsequently in Figs. \ref{fig:cvc_elgen} and \ref{fig:cvc_elgdist}. The cluster Sa--Sc data are also well fit by this line with an rms scatter of 0.18 compared to 0.31 for the Sa--Sc field line.

Figure \ref{fig:cvc_elgen} shows the cluster ELG sample split into enhanced and non-enhanced objects. Both subsamples have the majority of their points above the field line and are much better fit by the dotted ELG line. No significant difference is seen between the two samples, although again, the galaxies with the highest values of $C_{\rm{H}\alpha}$ are in the non-enhanced sample. 

More obvious are the results shown in Fig. \ref{fig:cvc_elgdist}, where the sample is split into disturbed and undisturbed objects. Although there is some scatter, the majority of disturbed objects lie above not only the field line ($\sim$75\%), but also the best fit line to the cluster as a whole (63\%). The undisturbed sample, on the other hand, is broadly consistent with the field line, with most objects (68\%) lying below the dotted best fit cluster line, although with a few high $C_{\rm{H}\alpha}$ exceptions. This is consistent with the idea that a number of objects may be experiencing an enhancement of star formation in circumnuclear regions due to tidal interactions. The relative scarcity of high $C_{\rm{H}\alpha}$ objects in the undisturbed ELG sample, compared to the Sa--Sc sample, may be because many of those objects with reduced star formation in their outer regions become too faint in \Ha\ to be detected by the OPS.

\section{Concentration indices and \Ha\ equivalent width}
\label{sec:cew}

\begin{figure}
\includegraphics[width=87mm]{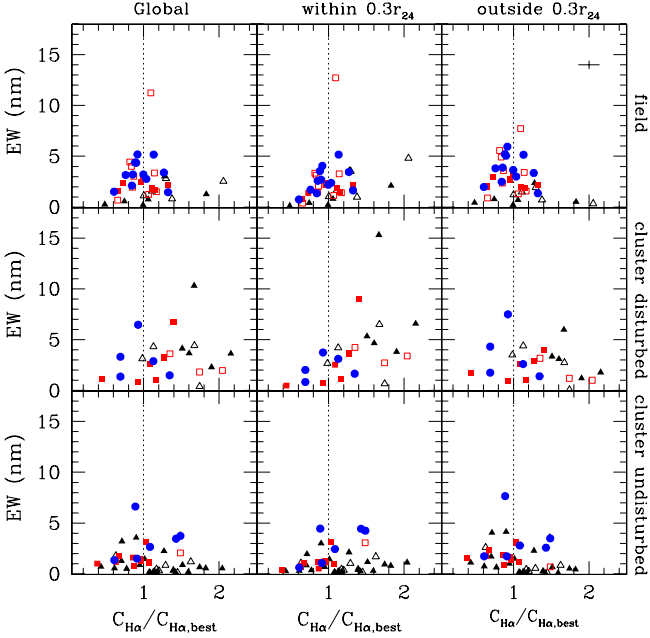}
\caption{Global EW (left), and EW within (centre) and beyond (right) 0.3$r_{24}$ plotted against $C_{\rm{H}\alpha}$/$C_{\rm{H}\alpha_{best fit}}$, for field (top) and cluster disturbed (centre) and undisturbed (bottom) Sa--Sc galaxies. EW errors are $\sim$5--15\% for high ($> 2$nm), 15--25\% for moderate (1--2 nm) and 25--100\% for low ($<$ 1 nm) EW objects.Typical error bars are shown in the upper-right corner of the plot, and apply to all nine frames.}
\label{fig:cew_sac}
\end{figure}

Figures \ref{fig:cvc_all}--\ref{fig:cvc_elgdist} have been interpreted here in terms of concentrated \Ha\ emission being due to enhanced star formation in the central regions in some galaxies and truncation of the star forming disk in others.  We will now compare the relative \Ha\ concentrations to central and global values of EW in each galaxy, in an attempt to gain further understanding of these processes. Fig. \ref{fig:cew_sac} shows (left to right) the global EW, the EW for the inner 30\% of each galaxy (within 0.3$r_{24}$), and the EW outside 0.3$r_{24}$ plotted against $C_{\rm{H}\alpha}$/$C_{\rm{H}\alpha_{best fit}}$ for the Sa--Sc sample. The cluster data are split into disturbed and undisturbed populations, and equivalent data for the field sample are also shown. The value on the $x$-axis, $C_{\rm{H}\alpha}$/$C_{\rm{H}\alpha_{best fit}}$, is the measured \Ha\ concentration index for each galaxy divided by the $C_{\rm{H}\alpha_{best fit}}$ value, calculated from the $R$-band concentration index of the galaxy using the best fit line equation, which has the form:

\begin{equation}
\nonumber
C_{\rm{H}\alpha_{best fit}}=1.269C_{30}-0.170.
\end{equation}

A value of $C_{\rm{H}\alpha}$/$C_{\rm{H}\alpha_{best fit}}$ equal to one indicates that the object lies on the best fit field lines in Figs. \ref{fig:cvc_all}--\ref{fig:cvc_cldist}, and this is shown as a dotted line in Fig. \ref{fig:cew_sac}. Any value above one means that the object has more centrally concentrated \Ha\ emission than would be expected from the $R$-band $C_{30}$ value, where the latter can be taken as a quantitative and unbiased proxy for galaxy type.

Figure \ref{fig:cew_sac} shows data for the complete Sa--Sc sample. The different points indicate morphological types, as used in previous figures. The top panels show that the field sample is tightly clustered about the best fit line, with the highest EW objects lying closest to the line. Even the bright object NGC 3310, which stands out as having enhanced emission, lies very close to the line, with relatively bright \Ha\ emission in global, inner and outer bins. One Sab field galaxy also stands out as having both concentrated emission and relatively enhanced EW for its type, within 0.3$r_{24}$, but normal to reduced emission in the outer regions. This object is NGC 2146 (UGC 3429), which is an infrared-luminous galaxy with a clearly disturbed morphology, thought to be a postmerger object undergoing a burst of star formation in its central regions (\citealt{hut90,mar01}). There is also an Sa object (UGC 7054) with relatively concentrated \Ha\ emission, which may be due to a slight enhancement of emission within 0.3$r_{24}$. Table \ref{tbl:cHacHabf} summarises the mean values of EW, with standard errors on the mean, for each of the samples shown in Fig.\ref{fig:cew_sac}, split into those with  $C_{\rm{H}\alpha}$/$C_{\rm{H}\alpha_{best fit}}$ ratios greater than or less than 1.0.

\begin{table*}
\begin{center}
\begin{small}
\begin{tabular}{|c||c|c|c|c|c|c|c|}
\hline
Sample   &  No. & Global  & Within 0.3$r_{24}$ & Outside 0.3$r_{24}$ \cr
\hline
Field & & & & \cr
$C_{\rm{H}\alpha}$/$C_{\rm{H}\alpha_{best fit}}<1$ & 24 & 2.25(0.32) & 1.57(0.25) & 2.74(0.36) \cr
$C_{\rm{H}\alpha}$/$C_{\rm{H}\alpha_{best fit}}>1$ & 20 & 2.56(0.52) & 2.85(0.59) & 2.14(0.40) \cr
\hline
Cluster undisturbed & & & & \cr 
$C_{\rm{H}\alpha}$/$C_{\rm{H}\alpha_{best fit}}<1$ & 18 & 1.72(0.35) & 1.15(0.25) & 2.14(0.40) \cr
$C_{\rm{H}\alpha}$/$C_{\rm{H}\alpha_{best fit}}>1$ & 28 & 1.14(0.23) & 1.34(0.25) & 0.95(0.22) \cr
\hline
Cluster disturbed & & & & \cr
$C_{\rm{H}\alpha}$/$C_{\rm{H}\alpha_{best fit}}<1$ & 6  & 2.71(0.87) & 1.74(0.53) & 3.29(0.98) \cr
$C_{\rm{H}\alpha}$/$C_{\rm{H}\alpha_{best fit}}>1$ & 17 & 3.45(0.56) & 4.62(0.84) & 2.50(0.36) \cr
\hline
ELG undisturbed & & & & \cr
$C_{\rm{H}\alpha}$/$C_{\rm{H}\alpha_{best fit}}<1$ & 12 & 2.53(0.50) & 1.75(0.35) & 3.10(0.58) \cr
$C_{\rm{H}\alpha}$/$C_{\rm{H}\alpha_{best fit}}>1$ & 19 & 2.23(0.36) & 2.57(0.41) & 1.86(0.35) \cr
\hline
ELG disturbed & & & & \cr
$C_{\rm{H}\alpha}$/$C_{\rm{H}\alpha_{best fit}}<1$ & 14 & 4.47(0.61) & 2.99(0.46) & 5.37(0.69) \cr
$C_{\rm{H}\alpha}$/$C_{\rm{H}\alpha_{best fit}}>1$ & 44 & 3.89(0.41) & 5.14(0.49) & 2.78(0.42) \cr
\hline
\end{tabular}
\caption{Mean EW values for galaxy subsets in each of the frames in Fig. \ref{fig:cew_sac},
 split into those with  $C_{\rm{H}\alpha}$/$C_{\rm{H}\alpha_{best fit}}$ ratios greater than or less than 1.0.}
\label{tbl:cHacHabf}
\end{small}
\end{center}
\end{table*}

The cluster objects in general are almost indistinguishable from the field sample for $C_{\rm{H}\alpha}$/$C_{\rm{H}\alpha_{best fit}} \lesssim 1$. However, there is a large spread in $C_{\rm{H}\alpha}$/$C_{\rm{H}\alpha_{best fit}}$ to the right of the dotted line. For the undisturbed objects this is largely due to a number of early-type objects with low emission across the galaxy, but particularly in the region beyond 0.3$r_{24}$. The disturbed objects, however, tend to have higher global EW values, and this enhancement seems to be strongest, for the majority of objects, within 0.3$r_{24}$. Of particular interest are the early type Sa--Sab spirals, which have global and inner EW values, and relative \Ha\ concentrations, well above most of their field counterparts. Many of these objects, however, also seem to have some enhancement of star formation in their outer disks. A similar plot for the complete ELG sample (not shown) suggests that this trend extends to earlier types and most of the peculiar objects (77\%) also have high \Ha\ concentrations and particularly large EW values within 0.3$r_{24}$. 


To summarise, the main result found here is that a large fraction of disturbed early-type cluster galaxies typically show higher \Ha\ EW values than similar galaxies from the control field or supercluster samples, and this strong emission is concentrated in the central regions of the affected galaxies.  

\section{Star formation within and outside the effective radius}
\label{sec:efr}

We now make a further test of the patterns of star formation found in cluster galaxies, making use of 
the effective radius, $r_{eff}$, of each galaxy. The effective radius is defined here as the semi-major axis of the ellipse that contains half of 
the total $R$-band light within the well-defined $r_{24}$ isophotal radius, thus avoiding difficulties in reliably determining the total luminosity which complicate the traditional definition of $r_{eff}$. By comparing the EW within $r_{eff}$ to the total EW within $r_{24}$ the relative central concentration of \Ha\ to $R$-band emission can be investigated. By definition, the ratio $F_{R_{eff}}$/$F_{R_{tot}}$, where $F_{R_{eff}}$ is the $R$-band flux density within $r_{eff}$ and $F_{R_{tot}}$ is the total $R$-band flux density within $r_{24}$, is 0.5. If the \Ha\ is distributed in the same way as the continuum light, then the ratio of \Ha\ flux within $r_{eff}$ to the total within $r_{24}$, $F_{\rm{H}\alpha_{eff}}$/$F_{\rm{H}\alpha_{tot}}$, is also 0.5, and the ratio $EW_{eff}/EW_{tot}$, given by:

\begin{equation}
\frac{EW_{eff}}{EW_{tot}}=\frac{F_{\rm{H}\alpha_{eff}}/F_{\rm{H}\alpha_{tot}}}{F_{R_{eff}}/F_{R_{tot}}}=\frac{F_{\rm{H}\alpha_{eff}}/F_{\rm{H}\alpha_{tot}}}{0.5},
\end{equation}

is equal to one. An object with highly concentrated \Ha\ emission, however, such that all star formation occurs within $r_{eff}$, will have $F_{\rm{H}\alpha_{eff}}$/$F_{\rm{H}\alpha_{tot}} = 1$ and $EW_{eff}/EW_{tot} = 2$. $EW_{eff}/EW_{tot}$ therefore ranges from 0--2 and gives a measure of how concentrated an object's star formation is compared to the underlying, older stellar population. 

\begin{figure}
\includegraphics[width=87mm]{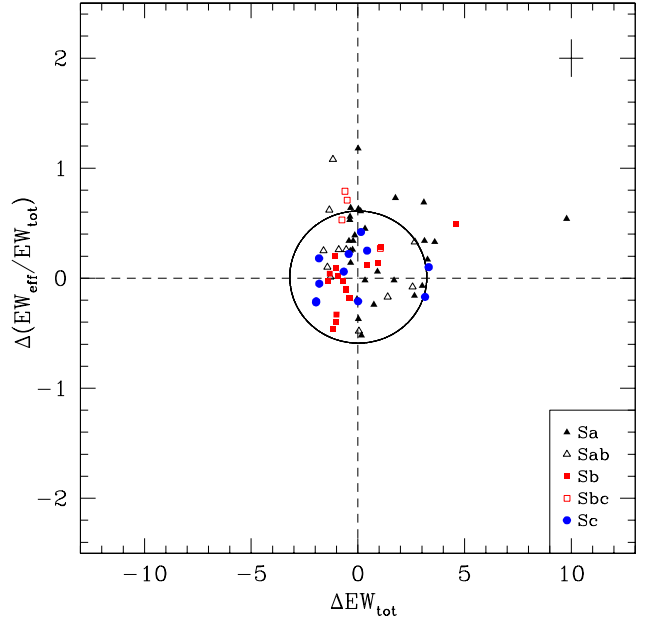}
\caption{Difference in $EW_{eff}/EW_{tot}$ and $EW_{tot}$ between cluster Sa--Sc galaxies and mean field values for a given type. The ellipse shows the 3$\sigma$ limits obtained from matching each field galaxy to the relevant field means. Typical error bars are shown in the upper-right corner of the figure.}
\label{fig:evdia}
\end{figure}

A test was carried out to identify the differences between the field and cluster populations. For each field type, a biweight estimator method was used to find the mean value and standard deviation of $EW_{tot}$ and $EW_{eff}/EW_{tot}$. These field values are listed in Table \ref{tbl:ewmns}, and set the reference values relative to which we calculate values for the cluster samples. The mean relative concentration of \Ha\ to $R$-band, represented here by $EW_{eff}/EW_{tot}$, again remains fairly constant across all field types, although the earliest types (S0$+$ and Sa) show a markedly larger dispersion than later types, and the mean for the S0$+$ galaxies is somewhat lower than the global mean, presumably reflecting the large bulges dominated by old stellar populations in many of these galaxies. Each cluster galaxy is then compared to the field mean for its type and the quantities $\Delta(EW_{eff}/EW_{tot})$ and $\Delta(EW_{tot})$ are found as the differences between the galaxy and mean field values. These are plotted for the Sa--Sc sample in Fig. \ref{fig:evdia}.

\begin{table*}
\begin{center}
\begin{small}
\begin{tabular}{|c||c|c|c|c|c|c|c|c|}
\hline
Type & S0+ & Sa & Sab & Sb & Sbc & Sc & Sc+ & all spirals \cr
\hline
\hline
$<EW_{tot}>$ & 0.93 & 0.54 & 1.74 & 2.16 & 2.57 & 3.31 & 3.20 & 2.65 \cr
$\sigma_{EW_{tot}}$ & 0.47 & 0.45 & 0.84 & 0.46 & 2.05 & 1.25 & 1.30 & 1.49 \cr
\hline
$<EW_{eff}/EW_{tot}>$ & 0.56 & 0.91 & 1.06 & 0.86 & 0.88 & 0.86 & 0.90 & 0.90 \cr
$\sigma_{EW_{eff}/EW_{tot}}$ & 0.62 & 0.55 & 0.14 & 0.18 & 0.16 & 0.14 & 0.30 & 0.25 \cr
\hline
\end{tabular}
\caption[Field means and standard deviations of $EW_{tot}$ and $EW_{eff}/EW_{tot}$]{Mean field values and standard deviations of $EW_{tot}$ and $EW_{eff}/EW_{tot}$.}
\label{tbl:ewmns}
\end{small}
\end{center}
\end{table*}

\begin{figure}
\includegraphics[width=87mm]{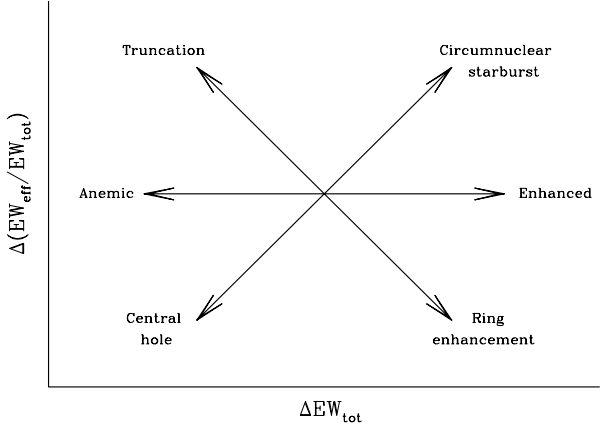}
\caption{Diagram showing a number of processes that occur in cluster galaxies and the effect each would have on the total \Ha\ EW and on its concentration relative to the continuum emission.}
\label{fig:evpos}
\end{figure}

If star formation were to be reduced equally across both the bulge and disk of a cluster galaxy, such that it became anemic, the value of $\Delta(EW_{tot})$ would be negative, but the concentration parameter $\Delta(EW_{eff}/EW_{tot})$ would remain constant, moving the galaxy towards the left in Fig. \ref{fig:evdia}. Truncation of SF in outer regions through gas stripping, on the other hand, would also reduce the total EW, but the concentration parameter would increase as most of the remaining star formation would be taking place in central regions. The galaxy would therefore move to the top left of Fig. \ref{fig:evdia}. Figure \ref{fig:evpos} shows for these, and a number of other possible processes that may occur in cluster galaxies, the direction in which each would move an object in Fig. \ref{fig:evdia}.

\begin{table*}
\begin{center}
\begin{small}
\begin{tabular}{|c||c|c|c|c|c|c|}
\hline
Type & Sa & Sab & Sb & Sbc & Sc & Sa - Sc \cr
\hline
$<\Delta EW_{tot}>$ & 1.14(0.41) & --0.15(0.49) & --0.23(0.40) & --0.20(0.42) & --0.14(0.57) & 0.35(0.23) \cr
$<\Delta(EW_{eff}/EW_{tot})>$ & 0.27(0.07) & 0.20(0.12) & --0.01(0.07) & 0.58(0.11) & 0.03(0.07) & 0.18(0.04) \cr
$<log({EW_{tot}/EW_{fmean}}>$ & 0.19(0.10) & --0.27(0.15) & --0.14(0.07) & --0.05(0.07) & --0.09(0.07) & --0.01(0.06) \cr
\hline
\end{tabular}
\caption{Mean values of $\Delta EW_{tot}$, $\Delta (EW_{eff}/EW_{tot})$ and log$(EW_{tot}/EW_{fmean})$ for cluster Sa - Sc galaxies.}
\label{tbl:dewSac}
\end{small}
\end{center}
\end{table*}

\begin{table*}
\begin{center}
\begin{small}
\begin{tabular}{|c||c|c|c|c|c|c|c|c|}
\hline
Type & S0+ & Sa & Sab & Sb & Sbc & Sc & Sc+ & all spirals \cr
\hline
$<\Delta EW_{tot}>$ & 2.18(0.66) & 2.55(0.70) & 0.80(0.68) & 0.34(0.48) & --0.65(0.06) & 0.53(0.65) & 3.50(0.50) & 1.42(0.26) \cr
$<\Delta(EW_{eff}/EW_{tot})>$ & 0.92(0.12) & 0.44(0.17) & 0.08(0.16) & 0.15(0.12) & 0.52(0.16) & 0.01(0.07) & --0.02(0.13) & 0.31(0.05) \cr
$<log({EW_{tot}/EW_{fmean}}>$ & 0.31(0.18) & 0.63(0.10) & 0.04(0.16) & --0.01(0.07) & --0.13(0.01) & 0.01(0.08) & 0.32(0.03) & 0.17(0.05) \cr
\hline
\end{tabular}
\caption{Mean values of $\Delta EW_{tot}$, $\Delta (EW_{eff}/EW_{tot})$ and log$(EW_{tot}/EW_{fmean})$ for the cluster ELG sample.}
\label{tbl:dewelg}
\end{small}
\end{center}
\end{table*}

For both the complete Sa--Sc and ELG samples the quantities $\Delta (EW_{tot})_{field}$ and $\Delta (EW_{eff}/EW_{tot})_{field}$ were also calculated by matching each field galaxy to the mean values for its type. These were then used to calculate the standard deviations in the field sample, and the locus of the 3$\sigma$ limits in each direction is indicated by the central circle in Fig. \ref{fig:evdia}.

From Fig. \ref{fig:evdia}, it can be seen that a majority of cluster galaxies have more concentrated \Ha\ emission than their mean field counterparts, with $\sim$62\% of points falling above the dashed horizontal line. There also appear to be more cluster objects with lower total EWs, with nearly 60\% of points lying to the left of the vertical dashed line. This is, however, dependent on type, with 21 of 30 late type Sb--Sc galaxies (70\%) having lower total EWs than the field means. The Sa--Sab galaxies, on the other hand, are approximately evenly split around the mean field line, although none appear in the bottom left quadrant (which would imply a central hole in the star formation) of the plot. Similarly, very few late type objects are seen in the bottom right quadrant (enhancement of star formation out in the disk). The majority of cluster objects lie within the field 3$\sigma$ ellipse, but there are a number of points lying beyond 3$\sigma$ in the top half of the plot, particularly in the top right hand quadrant, suggesting an enhancement of circumnuclear star formation in these objects. A few objects are also consistent with truncation, and the low mean EW values of the earlier types mean that some truncated early type spirals may be missed, so the numbers found in this category should be considered a lower limit.

The probability of a point falling outside the 3$\sigma$ ellipse was estimated assuming that the field population, from which the ellipse was calculated, is consistent with a binormal, or two-dimensional Gaussian, distribution such that the probability, $P(x,y)$, of a point falling at position $(x,y)$ is given by:

\begin{equation}
P(x,y)=\frac{1}{2\pi\sigma_{x}\sigma_{y}\sqrt{(1-\rho^{2})}}\times e^{\left\{-\frac{1}{2(1-\rho^{2})}\left[\left(\frac{x-\mu_{x}}{\sigma_{x}}\right)^{2}+\left(\frac{y-\mu_{y}}{\sigma_{y}}\right)^{2}-2\rho\left(\frac{x-\mu_{x}}{\sigma_{x}}\right)\left(\frac{y-\mu_{y}}{\sigma_{y}}\right)\right]\right\}},
\label{eq:prob}
\end{equation} 

\citep{bar89} where $\mu_{x}$ and $\mu_{y}$ are the means, and $\sigma_{x}$ and $\sigma_{y}$ the standard deviations in the $x$ and $y$ directions respectively, and $\rho$ is a correlation coefficient given by:

\begin{equation}
\rho=\frac{(\overline{xy})-\overline{x}. \overline{y}}{\sigma_{x}\sigma_{y}}.
\end{equation}

In this case $\Delta (EW_{tot})_{field}$ and $\Delta (EW_{eff}/EW_{tot})_{field}$ are uncorrelated, so $\rho\sim0$. Equation \ref{eq:prob} can therefore be simplified to:

\begin{equation}
P(x,y)=\frac{1}{2\pi\sigma_{x}\sigma_{y}}\times e^{\left\{-\frac{1}{2}\left[\left(\frac{x-\mu_{x}}{\sigma_{x}}\right)^{2}+\left(\frac{y-\mu_{y}}{\sigma_{y}}\right)^{2}\right]\right\}}.
\label{eq:prob2}
\end{equation} 

The equal probability contours are the ellipses for which the exponent in Equation \ref{eq:prob2} is constant, so the 3$\sigma$ ellipse used here can by described by:

\begin{equation}
\left(\frac{x-\mu_{x}}{\sigma_{x}}\right)^{2}+\left(\frac{y-\mu_{y}}{\sigma_{y}}\right)^{2} = 9.
\end{equation}

The probability of a point lying within this ellipse was estimated in two ways. Firstly a 500 $\times$ 500 grid was formed to cover the area of the ellipse and beyond. The probability of a point lying at the centre of each element was then found and this was assumed to be constant across the element, such that the probability of a point falling within each box could be calculated simply by multiplying the probability at the centre of the box by its area. The sum was then found of the probabilities associated with all elements whose centres lay within the 3$\sigma$ ellipse. This sum gives an estimate of the probability of a point falling within the 3$\sigma$ ellipse as 98.9\%.

The second method used 500 elliptical annuli, of increasing semi-major axis but constant width, instead of the grid. A probability was found of a point lying at the mean semi-major axis of each annulus, along the $x$-axis, and multiplied by its area, as before, to find the probability of a point falling within that annulus. The sum of these probabilities once again gives the overall probability of falling within the ellipse as 98.9\%. The probability of finding a point outside the 3$\sigma$ ellipse is therefore just 1.1\%.

Assuming a binomial distribution, the probability of finding 16 out of 69 points (23\%) in Fig. \ref{fig:evdia} outside the 3$\sigma$ ellipse, when expecting only 1.1\%, is less than $5\times10^{-17}$.

\begin{figure}
\includegraphics[width=87mm]{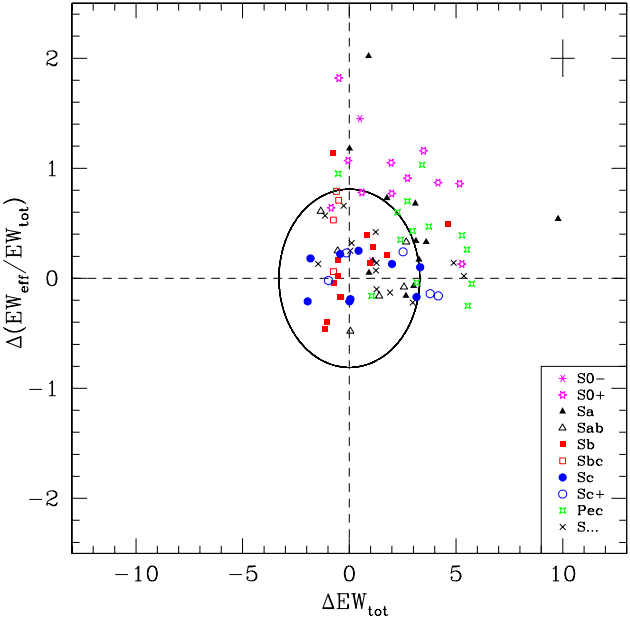}
\caption{As Fig. \ref{fig:evdia} but for the cluster ELG sample. Here all cluster galaxies of type E--S0/a are matched to the field S0/a means and the peculiar and unknown spiral galaxies are matched to mean field spiral values. Typical error bars are shown in the upper-right corner of the figure.}
\label{fig:evdia_elg}
\end{figure}

Figure \ref{fig:evdia_elg} shows a similar plot for the cluster ELG sample. Here the peculiar (Pec) and unknown spiral (S...) galaxies are compared to mean field spiral values, weighted according to the morphological make up of the cluster, whilst E--S0/a (S0$-$ and S0+) objects are compared to the mean values from the field S0/a sample. The one ELG cluster galaxy of unknown type is not included here. The ellipse corresponds to the 3$\sigma$ limits estimated from the field galaxy sample as explained above.  Once again the majority of cluster objects (72\%) have more concentrated \Ha\ emission than the field mean for their type. As expected the ELG sample galaxies also tend to have higher total EWs ($\sim$71\% to the right of the vertical dashed line) due to the effects of the sample selection, however, the cluster points would still be expected to lie largely within the field ellipse. For the ELG sample, some 40\% of points lie beyond the 3$\sigma$ ellipse. The probability of finding 36 of 89 points outside this ellipse by chance, assuming the same parent population as the field galaxy sample, is $\sim 2\times 10^{-46}$. 

Around half (51\%) of all ELG cluster galaxies fall in the top right quadrant of Fig. \ref{fig:evdia_elg}, suggestive of increased star formation in central regions. Of the 45 galaxies in this quadrant, 27 (60\%) lie beyond 3$\sigma$.

\begin{figure}
\includegraphics[width=87mm]{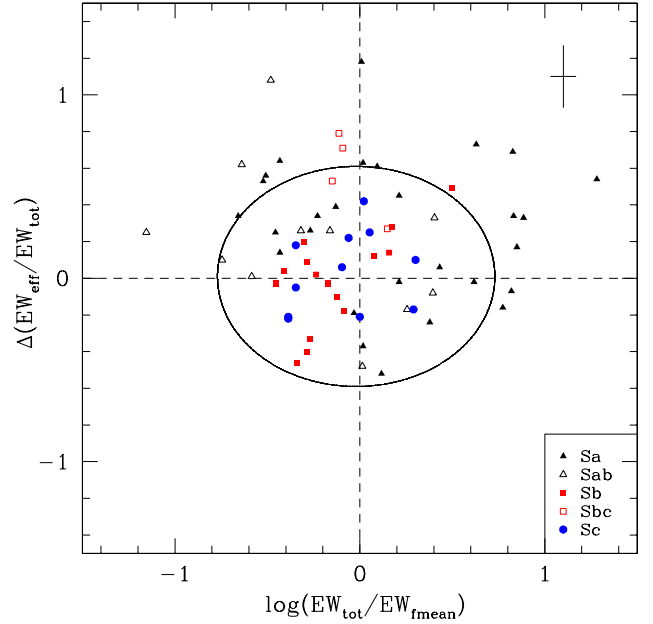}
\caption{Decimal logarithm of the total EW for each Sa--Sc cluster galaxy divided by mean field value for that type, plotted against $\Delta(EW_{eff}/EW_{tot})$ used previously. Twenty-one galaxies now lie beyond the 3$\sigma$ field ellipse. Typical error bars are shown in the upper-right corner of the figure.}
\label{fig:evdia_r}
\end{figure}

\begin{figure}
\includegraphics[width=87mm]{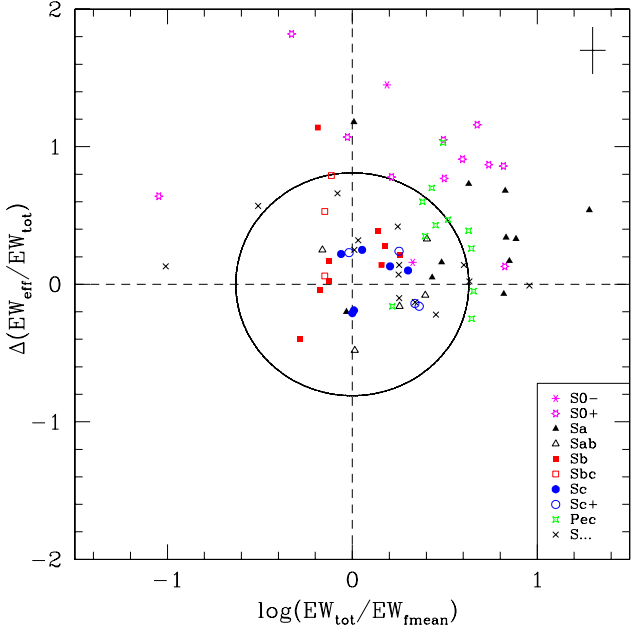}
\caption{As Fig. \ref{fig:evdia_r}, but here showing the cluster ELG sample.  Typical error bars are shown in the upper-right corner of the figure.}
\label{fig:evdia_relg}
\end{figure}

Although it is possible to identify most or all objects with enhanced circumnuclear emission in Figs. \ref{fig:evdia} and \ref{fig:evdia_elg}, a number of galaxies, in particular earlier types, with reduced star formation may be missed. This is because, due to the relatively low mean EWs for these types, even quite large fractional changes in SF activity can result in small absolute changes in EW. Figures \ref{fig:evdia_r} and \ref{fig:evdia_relg} therefore show the decimal logarithm of the ratio of each cluster galaxy's total EW to the field mean, plotted against the same concentration parameter. Again 3$\sigma$ ellipses are shown, generated from matching the field galaxies to the field means as before. Although the points falling in each quadrant remain the same, galaxies with highly reduced emission now also stand out beyond the field ellipse. 

Figure \ref{fig:evdia_r} shows the Sa--Sc cluster sample. Some 21 points (30\%, probability = $1.2 \times 10^{-24}$) now lie outside the 3$\sigma$ ellipse, compared to 16 in Fig. \ref{fig:evdia}, and 18 of these objects are types Sa--Sab. Nearly 43\% of cluster Sa and Sab galaxies in the Sa--Sc sample lie beyond 3$\sigma$, whilst over 90\% of later type galaxies remain within the field limits. Although 10 galaxies are still found beyond the ellipse in the top right hand quadrant, 9 objects are now also seen beyond 3$\sigma$ towards the top left of Fig. \ref{fig:evdia_r}, suggesting that, as well as a number of galaxies with enhanced circumnuclear emission, there is also a population of objects with truncated star forming disks. Two further Sa galaxies appear to be experiencing an enhancement of star formation further out in the disk, beyond the effective radius. The means and standard errors on the means of the quantities plotted in Figs. \ref{fig:evdia} and \ref{fig:evdia_r}, for the Sa - Sc cluster sample,  are summarised in Table \ref{tbl:dewSac}.

Figure \ref{fig:evdia_relg} shows a similar plot for the ELG sample, where again the one galaxy with no type information is excluded. Here the majority of points (71\%) beyond 3$\sigma$ still lie to the top right, although a few galaxies, particularly of later types or peculiar/S... galaxies, which previously appeared beyond 3$\sigma$, now lie within the field limits, probably due to the relatively high mean EWs for these objects. The galaxies with enhanced circumnuclear emission are again generally early types. A few galaxies also show signs of truncation, and we repeat that the number of these found here may be a lower limit, as galaxies with highly truncated disks, and therefore greatly reduced star formation, may not have been detected by the OPS, and would therefore not have been included in the ELG sample. The means and standard errors on the means of the quantities plotted in Figs. \ref{fig:evdia_elg} and \ref{fig:evdia_relg}, for the ELG cluster sample,  are summarised in Table \ref{tbl:dewelg}.

\section{Discussion}

This paper has looked at the radial distributions of $R$-band and \Ha\ light in samples of cluster and field galaxies using concentration indices and other statistical measures.  In this section we will summarise the main results, compare them with the results of related studies from  the literature, and draw conclusions about the effect of the cluster environment on star formation within disk galaxies.

The analysis initially addressed the $R$-band $C_{30}$ concentration index, which relates to bulge-to-disk ratio, and is therefore well correlated with morphological type, at least for the field and supercluster samples. The first finding of this paper is that $C_{30}$ is much more weakly correlated with type for the cluster galaxies; these show a much wider range in $C_{30}$ than the field galaxies for a given type. In particular, half of all cluster Sa galaxies have $C_{30}$ values below that for any field Sa. The correlation between $C_{30}$ and type for the field sample also extends to \Ha\ EW, with lower concentration, later-type spirals tending to have higher EW values, as has been shown by many previous studies \citep{ken83a,ken83b,deh94,ryd94,you96,ken98}. However the correlation between $C_{30}$ and EW is stronger for the cluster sample than the weak dependence of $C_{30}$ on type shown by the same galaxies.

The \Ha\ concentration parameter $C_{\rm{H}\alpha}$ also correlates well with $T$-type for the field galaxies, but this trend is found to be much weaker for the cluster galaxies. There is little correlation between $C_{\rm{H}\alpha}$ and \Ha\ EW for field, supercluster or cluster galaxies.

In Section \ref{sec:cvc}, the \Ha\ and $R$-band concentration indices were compared for the field and the Sa--Sc and ELG cluster samples. Nearly a half of the Sa--Sc cluster galaxies have more than 50\% of their star formation within 0.3$r_{24}$, compared to less than a quarter of field and supercluster objects. The $C_{\rm{H}\alpha}$ distributions of the field and cluster Sa--Sc samples are different at $>$98\% significance. Comparing to a best fit field line, cluster galaxies are also found to have relatively more concentrated \Ha\ emission than would be expected from their $C_{30}$ values. This high concentration is particularly marked for disturbed galaxies. This suggests that a number of objects may be experiencing an increase of star formation in central regions, and in light of the later analysis presented in Section \ref{sec:efr} this appears to be the dominant effect for the present sample of galaxies.  However, it should also be noted that a smaller number of non-enhanced and undisturbed cluster galaxies also lie well above the best fit field line, and may instead be experiencing a reduction of star formation in their outer regions.

In both Sa--Sc and ELG samples the majority of galaxies showing \Ha\ emission that is more centrally concentrated than expected tend to be earlier types, although many of the peculiar galaxies in the ELG sample also fit into this category. The idea that such concentrated \Ha\ emission is caused by an increase of star formation in the central regions of some objects is backed up in Section \ref{sec:cew} where many objects with relatively high $C_{\rm{H}\alpha}$ values show particularly high EWs within 0.3$r_{24}$. There does, however, also seem to be a population of high \Ha\ concentration galaxies with low overall EWs, i.e. in these galaxies SF is suppressed everywhere but particularly strongly in their outer regions.

Section \ref{sec:efr} contains a comparison of total EW and $EW_{eff}/EW_{tot}$, where the latter represents the relative concentration of \Ha\ to $R$-band emission, for each cluster galaxy, to the field means for galaxies of the same type. This analysis shows that most cluster galaxies lie within the parameter range expected for `normal' galaxies as defined by the field population, but that the fraction of cluster galaxies with `abnormal' SF patterns is far more than should be expected by chance.  Where this abnormal SF occurs, it tends to take the form of an overall increase in the luminosity-normalised SF activity, and this additional SF is concentrated in the central regions of the affected galaxies.

We will now compare these conclusions with other studies in the literature that have addressed similar properties of star-forming galaxies. In our discussion of H$\alpha$ and $R$-band concentration indices in section \ref{sec:cvc}, Figs. 5 to 9 show that generally the cluster galaxies have more centrally concentrated SF as revealed by \Ha\ than would be be expected for field galaxies with the same $R$-band light distribution, the latter being indicated by the $C_{30}$ index.  There are several possible explanations for this, but a very plausible explanation is that it is associated with enhanced SF that is concentrated in the nuclear regions of the galaxies concerned. In support of this, it has been long established that interacting and disturbed galaxies show enhanced SF overall \citep{lar78,jos84,lon84,ken87}. This SF has been shown to occur preferentially in central regions of the affected galaxies, both in observational studies \citep{ken84a,kee85}, and in theoretical simulations of the gas response to interactions \citep{mih92,ion04}. This may be the explanation of the centrally concentrated emission in the cluster galaxy population, particularly for the disturbed subsample.
However, these plots are also consistent with truncation of SF in outer disk regions following removal of outlying disk gas, the preferred interpretation of the detailed Virgo cluster study of \citet{koo04b,koo04a}.

In order to break this degeneracy, and to test other possible characteristic 
modes of SF in these cluster galaxies, the plot used to construct Figs. \ref{fig:evdia}, \ref{fig:evdia_elg}, \ref{fig:evdia_r} and \ref{fig:evdia_relg}, and interpreted in Fig.~\ref{fig:evpos}, was introduced.  These figures show that while there are some galaxies showing the truncated pattern of SF favoured by \citet{koo04b,koo04a} and \citet{dal01}, a larger fraction of the galaxies showing unusual patterns of SF have centrally enhanced SF, i.e. both their central EW and global EW values are unusually high for their type.

The discussion of the patterns of SF observed in cluster galaxies has centred so far on the highly unusual outliers that are found beyond the 3~$\sigma$ limits defined by the field galaxies.  However, close examination of Figs.~\ref{fig:evdia_elg} and \ref{fig:evdia_relg} shows that even within the 3~$\sigma$ limits, the cluster galaxies are not randomly distributed, and tend to favour the upper-right quadrant.  If this distribution is interpreted in terms of central starburst activity, this indicates that such activity can occur with a range of intensities.  There is no evidence of a threshold, above which intense activity is induced, but rather of a continuum of central SF activity strengths.

One pattern that is not observed in the present sample is apparently `anemic' systems \citep{van91,elm02}, with globally-suppressed gas-mass fractions and SF uniformly reduced across disk and central regions.  This may mean that such systems do not exist at all in the clusters studied here, or that the time in which they would be observed in such a state is short, before their overall emission-line activity sinks to unobservable levels and they become observationally passive galaxies. 

No cluster galaxies are seen with SF that is relatively stronger in their outer than in their central regions.  Such a pattern might be expected in galaxies with strong outer rings of SF, as found for some field galaxies by \citet{hod83}, which would shift points to the lower-right quadrant of Figs.~\ref{fig:evdia} - \ref{fig:evdia_relg}.   Alternatively, galaxies that have formed large bulges with suppressed SF while retaining SF in their outer disks might be expected in the lower-right quadrant.  However, we repeat that no obvious examples of either type are seen in the present cluster sample.

One final discussion point is to recall the conclusion from \citet{bre10} that the disturbed, high-EW galaxies show some evidence for larger-than-average velocities within their host clusters.  This may indicate that these galaxies are not virialised, and may be largely an infalling population.  This is consistent with the inferred processes affecting their star formation being relatively short-lived and one-off occurrences such as interaction-driven starbursts.  Other processes, such as `strangulation' and `harassment', might be occurring but over longer timescales, resulting in more virialised populations of galaxies with suppressed SF, that would not have been efficiently picked up by the OPS.

\section{Conclusions}

The primary results and conclusions from this paper are as follows:

\begin{itemize}
\item The $R$-band concentration parameter, $C_{30}$, correlates well with morphological type for the field sample, but this correlation is much weaker for the cluster data, with, for example,  half of all cluster Sa galaxies having $C_{30}$ values below that for any field Sa spiral (Fig. \ref{fig:C30dept}).
\item \Ha\ EW is found to be correlated with  $C_{30}$ for all samples
(Fig. \ref{fig:ewdepc30}). This correlation is strongest for the cluster data.
\item In Figs. \ref{fig:cvc_all}--\ref{fig:cvc_elgdist}, cluster galaxies are found typically to have higher $C_{\rm{H}\alpha}$ values than their field counterparts, and more concentrated \Ha\ emission than would be expected from their $C_{30}$ values. The $C_{\rm{H}\alpha}$ distributions of field and cluster Sa--Sc galaxies are different at $>$98\% significance.
\item The majority of galaxies showing the most concentrated \Ha\ emission, relative to $C_{30}$, are early-type spirals and peculiar galaxies. They are also more likely to be tidally disturbed.
\item Figure \ref{fig:cew_sac} shows that for the disturbed population these high $C_{\rm{H}\alpha}$ values are largely due to higher levels of star formation in central galaxy regions. However, there are also a number of high \Ha\ concentration, mainly undisturbed, galaxies with low EW values, particularly in their outer regions.
\item Figures \ref{fig:evdia} and \ref{fig:evdia_elg} confirm the result that some cluster galaxies, mainly disturbed peculiar and early spiral types, are experiencing an enhancement of star formation in their circumnuclear regions.
\item A population of galaxies with truncated disk star formation is also identified (Figs. \ref{fig:evdia_r} and \ref{fig:evdia_relg}). Many objects may be experiencing both an enhancement of circumnuclear star formation and a truncation of the star forming disk.
\end{itemize}

\begin{acknowledgements}
Based on observations made with the Nordic Optical Telescope, operated
on the island of La Palma jointly by Denmark, Finland, Iceland,
Norway, and Sweden, in the Spanish Observatorio del Roque de los
Muchachos of the Instituto de Astrofisica de Canarias.  The Jacobus
Kapteyn Telescope was operated on the island of La Palma by the Isaac
Newton Group in the Spanish Observatorio del Roque de los Muchachos of
the Instituto de Astrof\'isica de Canarias.  This research has made
use of the NASA/IPAC Extragalactic Database (NED) which is operated by
the Jet Propulsion Laboratory, California Institute of Technology,
under contract with the National Aeronautics and Space Administration.
We thank the referee for many helpful comments.
\end{acknowledgements}

\bibliographystyle{bibtex/aa}
\bibliography{refs}


\end{document}